\documentclass[12pt, a4paper]{iopart}
\usepackage{graphicx}
\usepackage{setstack}
\usepackage{amssymb}

\begin{document}
\title[Exchange interaction and Curie temperatures in tetrametal nitrides]{Exchange interactions and Curie temperatures of the tetrametal nitrides Cr$_4$N, Mn$_4$N, Fe$_4$N, Co$_4$N, and Ni$_4$N}
\author{Markus Meinert}
\address{Center for Spinelectronic Materials and Devices, Bielefeld University,
33501 Bielefeld, Germany}
\ead{meinert@physik.uni-bielefeld.de}

\date{\today}

\begin{abstract}
The exchange interactions of Cr$_4$N, Mn$_4$N, Fe$_4$N, Co$_4$N, and Ni$_4$N compounds with perovskite structure were calculated to obtain the Curie temperatures for these compounds from Monte Carlo calculations. Contrary to na\"{\i}ve expectation, the exchange interactions vary markedly among these five compounds. In Fe$_4$N,  the intra-sublattice interaction of the Fe $3c$ atoms is strongly negative, leading to a significant reduction of the Curie temperature. The calculated Curie temperatures are 291\,K (Cr$_4$N), 710\,K (Mn$_4$N), 668\,K (Fe$_4$N), 827\,K (Co$_4$N), and 121\,K (Ni$_4$N), in good agreement with experimental observations where available. The much lower Curie temperature of Ni$_4$N compared to fcc Ni is explained on the basis of the exchange interactions.

\end{abstract}

\maketitle

\section{Introduction}

Transition metal (TM) nitrides are well-known for their excellent properties as hard-coatings or diffusion barriers (TiN, ZrN, CrN, TaN$_x$) \cite{Holleck1986}. The useful magnetic properties of some magnetic nitrides have come only recently into the focus of research, particularly in the spintronics community, despite first experiments on iron nitrides date back to the 1940s \cite{Guillaud1946}.

Epitaxial magnetic tunnel junctions containing $\gamma'$-Fe$_4$N with the perovskite structure as an electrode have shown large negative TMR \cite{Komasaki2009} and inverse spin transfer torque (STT) switching \cite{Isogami2010}. Exchange bias stacks with $\gamma'$-Fe$_4$N and  rocksalt CoN as the antiferromagnet show negative exchange bias. Very large exchange bias at room temperature was observed in stacks with body-centered tetragonal MnN as the antiferromagnet \cite{Meinert2015}. Ferrimagnetic Mn$_4$N has a low magnetic moment and a high Curie temperature \cite{Mekata1962} and exhibits perpendicular magnetization on some substrates \cite{Yasutomi2014}, making it an ideal candidate for an electrode material in STT switching devices.

\begin{figure}[t]
\centering\includegraphics[width=6cm]{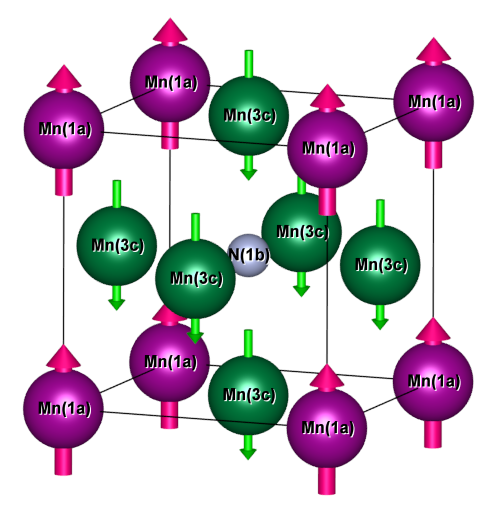}
\caption{\label{fig:Mn4N_structure} Structure and magnetic configuration of Mn$_4$N. The Wyckoff positions displayed in the figure are used to distinguish the inequivalent transition metal atoms.}
\end{figure}

Fe$_4$N and Mn$_4$N crystallize in the perovskite structure (space group $Pm\bar{3}m$, No. 221). The transition metals occupy the Wyckoff $1a$ (cube corner position $(0,0,0)$) and $3c$ (cube face $(1/2,1/2,0)$) positions, nitrogen sits on the Wyckoff $1b$ (cube center $(1/2,1/2,1/2)$) position. The structure and the collinear magnetic configuration of Mn$_4$N are depicted in Fig. \ref{fig:Mn4N_structure}. Other possible magnetic TM nitrides with this structure are Cr$_4$N, Co$_4$N, Ni$_4$N, and their alloys.

Because of thermal instability of the late transition metal nitrides in general, it is a challenging task to prepare samples of these compounds with ideal stoichiometry and chemical order. Thus, it is difficult to assess what the intrinsic properties of the ideal compounds are. A prominent and well studied example is Fe$_4$N, for which total magnetic moments between 8.7 and 11.6$\mu_\mathrm{B}$\,/\,cell were reported \cite{Diao1999, Frazer1958, Shirane1962, Ito2011, Atiq2008}. Even more intriguing is the case of giant magnetic moments in $\alpha''$-Fe$_{16}$N$_2$ films, which depend strongly on the precise film deposition conditions \cite{Takahashi2000}. Also Co$_4$N is a difficult case, as it decomposes at about 540\,K and loses nitrogen already at lower temperature \cite{Oda1987}, making it particularly difficult to obtain crystalline samples of correct stoichiometry. As a consequence, lattice constants between 3.59 and 3.74 \AA{} \cite{Oda1987, Terao60} and total magnetic moments ranging from 5.9 to 7.5 $\mu_\mathrm{B}$\,/\,cell were reported \cite{Ito2011_2, Silva2014, Gupta2015}, depending on the preparation conditions.

Many articles discussing the magnetic ground states, possible metamagnetic transitions under pressure, as well as densities of states of the tetrametal nitrides with perovskite structure are found in the literature \cite{Matar1988, Sakuma1991, Kuhnen2000, Patwari2001, Kokado2006, Matar2007, Houari2010, Imai2010, Takahashi2011, Hemzalova2013, Fang2014}. It is the aim of the present article to create an understanding for the experimentally observed Curie temperatures of the magnetic tetrametal nitrides, to predict the Curie temperatures for compounds for which a direct measurement is inaccessible, and to serve as a reference for the expected intrinsic properties of these compounds. To this end, first principles calculations were carried out and Heisenberg exchange interaction parameters were extracted. These were used to compute the Curie temperatures of the compounds based on the Heisenberg Hamiltonian.

\section{Computational approach}

The calculations presented in this work were done in the framework of density functional theory (DFT). In a first step, the lattice constants $a$ and magnetic configurations were determined with high-accuracy full-potential linearized augmented plane wave (FLAPW) calculations with the \textsc{elk} code \cite{elk}. Tight numerical parameters were chosen\footnote{APW+lo+LO basis set with optimized linearization energies, maximized muffin-tin radii, angular momentum cutoffs for wavefunctions $\ell_\mathrm{APW} = 10$ and for the potential $\ell_\mathrm{V} = 9$, plane wave expansion parameter for the wave functions $r_\mathrm{MT} G_\mathrm{max} = 8.0$ and charge density and potential expansion $G_\mathrm{max} = 15.0$\,a.u.$^{-1}$,  $22^3$ k-point mesh, smearing width 0.001\,Ha.} to ensure that the results are precise. In the second step, these lattice constants were used to compute the exchange interactions in a real space approach based on the Korringa-Kohn-Rostoker (KKR) multiple scattering theory. These calculations were performed with the spin-polarized relativistic KKR package \textit{Munich} SPR-KKR \cite{sprkkr}. The calculations were carried out in the full-potential mode, which is mandatory to reproduce the FLAPW results. In this mode, the unit cell is partitioned into non-overlapping polyhedra with the Wigner-Seitz method so that no interstitial regions occur. Empty polyhedra were introduced in the octahedral interstices (Wyckoff $3d$ positions $(0,0,1/2)$). An angular momentum cutoff of $\ell_\mathrm{max} = 2$ was used for the expansion of the Green function and the Brillouin zone was sampled with a $28 \times 28 \times 28$ $\bi{k}$-point mesh. The energy integration of the Green's function was done with 40 points along an arc in the complex plane. In order to improve the total charge convergence with respect to $l_\mathrm{max}$, Lloyd's formula was applied for the determination of the Fermi energy \cite{Lloyd72, Zeller08}. The exchange-correlation potential was modeled within the generalized gradient approximation of Perdew, Burke, and Ernzerhof \cite{PBE}. All calculations were carried out in the scalar-relativistic representation of the valence states, thus neglecting the spin-orbit coupling.

In the classical Heisenberg model the Hamiltonian of a spin system is given by
\begin{equation}\label{eq:heisenberg}
H = -\sum\limits_{i \neq j} J_{ij} \bi{e}_i \bi{e}_j ,
\end{equation}
with the Heisenberg pair exchange coupling parameters $J_{ij}$, and unit vectors $\bi{e}_{i}$ pointing in the direction of the magnetic moment on site $i$. SPR-KKR allows to calculate the exchange coupling parameters by mapping the full system onto a Heisenberg Hamiltonian. The parameters are determined within a real-space approach using the formula by Liechtenstein \textit{et al.} \cite{Liechtenstein87}. Positive sign of the $J_{ij}$ means a ferromagnetic interaction, negative sign means an antiferromagnetic interaction.

Approximate Curie temperatures were calculated within the mean field approximation (MFA). In a multi-sublattice system one has to solve the coupled equations
\begin{eqnarray}\label{eq:mfa}
\frac{3}{2} k_\mathrm{B} T_\mathrm{C}^{\mathrm{MFA}} \left< e^\mu \right> &= \sum\limits_\nu J_0^{\mu \nu} \left< e^\nu \right>\\
\mathrm{with}\quad J_0^{\mu \nu} &= \sum\limits_{\bi{r}\neq 0} J^{\mu \nu}_{0 \bi{r}}, \nonumber
\end{eqnarray}
where $\left< e^\nu \right>$ is the average $z$ component of the unit vector $e_\bi{r}^\nu$ pointing in the direction of the magnetic moment at site ($\nu$, $\bi{r}$). The coupled equations can be understood as an eigenvalue problem, where the largest Eigenvalue of the $J_0^{\mu \nu}$ matrix determines the Curie temperature \cite{Sasioglu05_1, Anderson63}. The $\bi{r}$-summation in Eq. (\ref{eq:mfa}) was taken to a radius of $r / a = 6.0$. It is well known that the MFA overestimates the Curie temperature \cite{Garanin96}, as long as the classical Heisenberg model is applicable, i.e. the absolute value of the local magnetic moments does not change upon rotation or spin wave excitation.

More accurate Curie temperatures were calculated with Monte Carlo (MC) simulations in the \textsc{vampire} atomistic spin dynamics program \cite{Evans14}. A $20\times 20 \times 20$ simulation supercell with periodic boundary conditions and 32,000 atoms was used, where nitrogen atoms and empty polyhedra were neglected for faster calculations. Because the exchange interactions have an RKKY contribution, they are oscillatory and decay with distance $r$ approximately as $J_{ij}(r) \sim 1/r^{3}$, whereas the number of interacting moments between $r$ and $r+\mathrm{d}r$ scales as $r^2 \mathrm{d}r$. Therefore, interactions up to $r/a = 6.0$ were included in the MC simulations to correctly reproduce the DFT ground state, resulting in a total of 14,352 pair interactions. 10,000 MC steps were done for thermalisation and measurement, respectively, at every temperature for temperatures in steps of approximately $T_\mathrm{C}^\mathrm{MFA} / 50$. The Curie temperatures were determined by interpolating the temperature dependence of the magnetization with cubic splines and finding the inflection point.

\Table{\label{overview} Lattice constants, magnetic moments and MFA Curie temperatures of the five magnetic TM$_4$N compounds. Magnetic moments are given in $\mu_\mathrm{B}$ per cell and temperatures are in Kelvin. The direction of the TM(1a) moment is defined as positive direction.}
\br
			& $a$ (\AA{})	 	&	$m^\mathrm{FLAPW}$		& $m^\mathrm{KKR}$	& $T_\mathrm{C}^\mathrm{MFA}$	& $T_\mathrm{C}^\mathrm{MC}$	&	$a_\mathrm{exp}$ (\AA{})	& $m_\mathrm{exp}$ 	& $T_\mathrm{C}^\mathrm{exp}$\\ \mr
Cr$_4$N	 	& $3.810$		&	$-1.06$		& $-0.77$		& $379$		&	291	& \textendash	&	\textendash	&	\textendash\\
Mn$_4$N 	 	& $3.742$		&	$1.44$		& $1.50$		& $870$		&	710	&	3.87$^\mathrm{a}$	&	1.1$^\mathrm{a}$	&	730$^\mathrm{a}$\\
Fe$_4$N 	 	& $3.784$		&	$9.86$		& $9.67$		& $995$		&	668	&	3.80$^\mathrm{d}$	&	9.1$^\mathrm{d}$	&	767$^\mathrm{d}$\\
Co$_4$N 	 	& $3.722$		&	$6.32$		& $6.22$		& $1025$		&	827	&	3.70$^\mathrm{c}$	&	7.4$^\mathrm{c}$	&	\textendash\\
Ni$_4$N 	 	& $3.732$		&	$1.57$		& $1.26$		& $143$		&	121	&	3.73$^\mathrm{b}$	&	1.6$^\mathrm{b}$	&	125$^\mathrm{b}$\\
\br
\end{tabular}
\item[$^\mathrm{a}$] Reference \cite{Mekata1962}
\item[$^\mathrm{b}$] extrapolated from data on (Fe$_{1-x}$Ni$_x$)$_4$N in Reference \cite{Diao1999}
\item[$^\mathrm{c}$] Reference \cite{Gupta2015}
\item[$^\mathrm{d}$] Reference \cite{Shirane1962}
\end{indented}
\end{table}

\Table{\label{moments} Local magnetic moments computed with an FLAPW program, with a KKR program, and obtained from neutron diffraction where available.}
\br
			& $m^\mathrm{FLAPW}_{1a}$	&	$m^\mathrm{FLAPW}_{3c}$	&	 $m^\mathrm{KKR}_{1a}$	&	$m^\mathrm{KKR}_{3c}$	&	$m^\mathrm{exp}_{1a}$		&	$m^\mathrm{exp}_{3c}$	\\ \mr
Cr$_4$N 	& 	1.79	&	$-$0.96	&	1.60	&	$-$0.79	&	\textendash	&	\textendash	 	\\
Mn$_4$N 	&	3.07	&	$-$0.64	&	3.04	&	$-$0.59	&	3.88$^\mathrm{a}$	&	-0.9$^\mathrm{a}$	 	\\
Fe$_4$N 	&	2.91	&	2.29	&	2.91	&	2.21	&	2.98$^\mathrm{b}$	&	2.01$^\mathrm{b}$	 	\\
Co$_4$N 	&	1.93	&	1.47	&	1.86	&	1.44	&	\textendash	&	\textendash	 	\\
Ni$_4$N 	&	0.68	&	0.30	&	0.60	&	0.22	&	\textendash	&	\textendash	 	\\
\br
\end{tabular}
\item[$^\mathrm{a}$] Reference \cite{Mekata1966}
\item[$^\mathrm{b}$] Reference \cite{Frazer1958}
\end{indented}
\end{table}

\section{Results}
\begin{figure*}[t]
\centering\includegraphics[width=\textwidth]{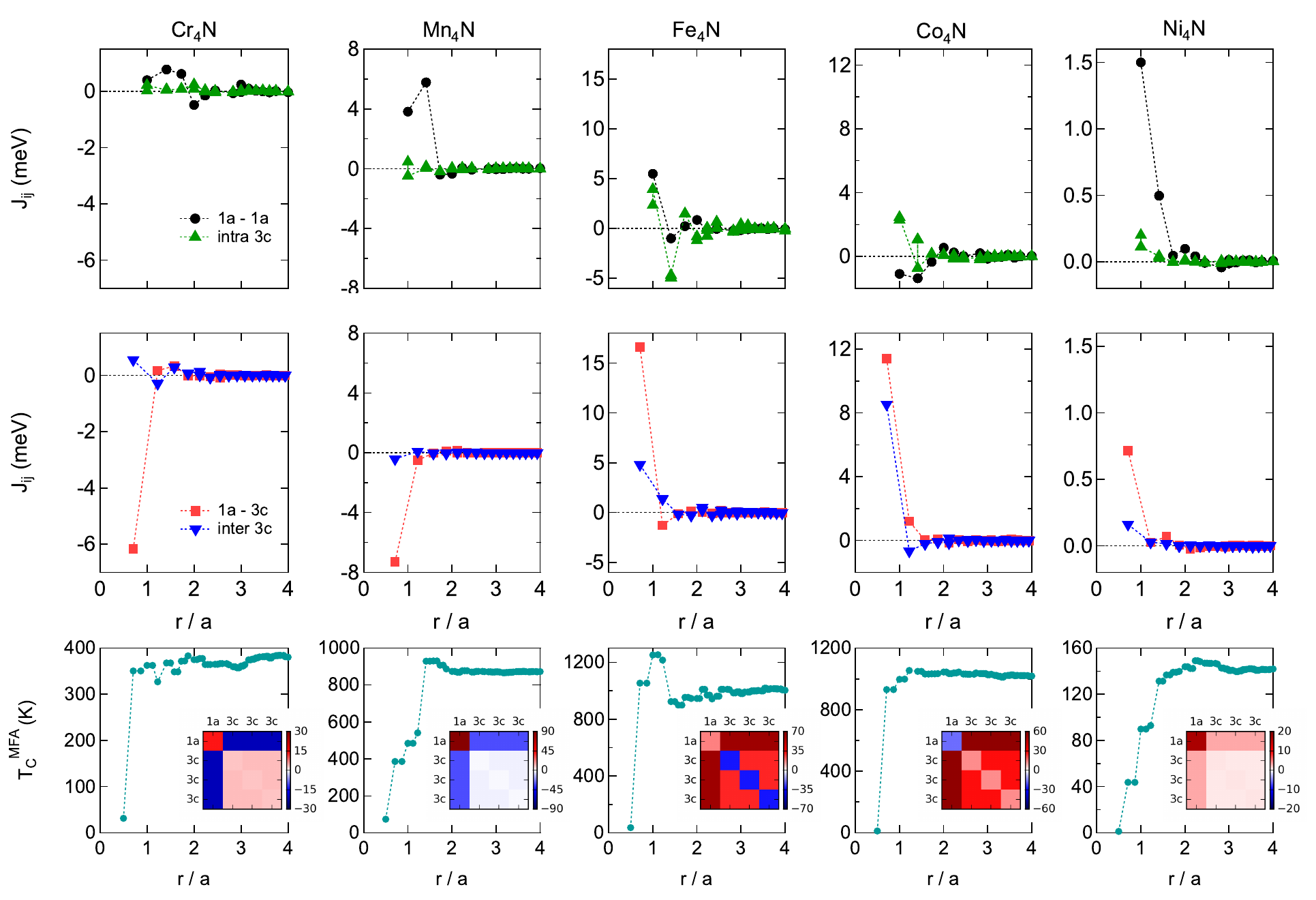}
\caption{\label{fig:Jij_results}Top row and middle row: Heisenberg exchange parameters as a function of interatomic distance in the tetrametal nitrides for intra-sublattice and inter-sublattice interactions, respectively. Bottom row: Mean-field estimate of the Curie temperature as a function of interatomic distances included in the construction of $J_0^{\mu\nu}$. Insets: Graphical representation of the coupling matrices $J_0^{\mu \nu}$. Note the scale changes between different compounds.}
\end{figure*}

 The calculated lattice constants, magnetic moments and Curie temperatures are collected in Table \ref{overview}. Additionally, the site-resolved magnetic moments are collected in Table \ref{moments}. The site- and distance-resolved exchange interactions, distance-resolved MFA Curie temperatures and coupling matrices $J_0^{\mu\nu}$ are presented in Figure \ref{fig:Jij_results}.
 
\subsection{Lattice constants and magnetic moments}
To the best of the author's knowledge, no experiments on Cr$_4$N were reported to date, so all results on Cr$_4$N are predictions.

For Mn$_4$N, the experimental lattice constant is 3.4\% larger than the theoretical value. Given the good agreement between calculated and experimental lattice constants for the other three compounds Fe$_4$N, Co$_4$N, and Ni$_4$N, this result could indicate that Mn$_4$N incorporates additional N atoms on the Wyckoff $3d$ positions, giving rise to an enhanced lattice constant. On the other hand, the calculated total moment is too large and the local magnetic moments are quite underestimated with respect to neutron diffraction results, see Table \ref{moments}. Both underestimations indicate that electron-electron correlation might play a significant role in Mn$_4$N and has to be taken into account for a correct description of this compound. The noncollinear magnetization components reported by Fruchart \textit{et al.} \cite{Fruchart1979} and Uhl \textit{et al.} \cite{Uhl1997} are rather small and are not expected to significantly increase the lattice constant.

The calculated lattice constant of Fe$_4$N is very close to the observed value and the calculated magnetic moment lies well within the reported range of magnetic moments between 8.7 and 11.6$\mu_\mathrm{B}$\,/\,cell \cite{Diao1999, Frazer1958, Shirane1962, Ito2011, Atiq2008}.  It was put forward by Blanc\'a \textit{et al.} that Fe$_4$N is found at a steep transition between low-spin and high-spin behaviour, which makes the magnetism of this compound particularly sensitive to the lattice constant. The same authors also pointed out that electron-electron correlation plays a significant role in Fe$_4$N. Notably, the local magnetic moments on both the $1a$ and $3c$ sites are higher than in bcc Fe ($m = 2.22\,\mu_\mathrm{B}$). The moments are however close to Fe in fcc crystals, such as pure fcc Fe with expanded lattice constant or FeNi alloys ($m_\mathrm{Fe} = 2.6\,\mu_\mathrm{B}$) \cite{Abrikosov2007}.

As mentioned in the introduction, Co$_4$N is a challenging case from the experimental point of view, because it is difficult to prepare the compound with the correct stoichiometry. With the lattice constant close to the theoretical value, the observed magnetic moment is significantly higher than the theoretical value. However, the value cited in Table \ref{overview} was obtained indirectly by polarized neutron reflectometry and might suffer from systematic overestimation. The theoretical average moment is slightly smaller than the value for hcp Co ($m = 1.72\,\mu_\mathrm{B}$). However, the $1a$ moment is larger in Co$_4$N, whereas the 3c moments are reduced with respect to the hcp Co value.

The calculated lattice constant and magnetic moment agree very well with extrapolated experimental results for Ni$_4$N. However, the results appear surprising at a first glance, in view of the fact that Ni$_4$N is very similar to fcc Ni, but has larger lattice constant. Due to the lattice expansion, one might expect \textit{higher} magnetic moment instead of the actually \textit{reduced} value as compared to fcc Ni ($m = 0.61\,\mu_\mathrm{B}$). Inspecting the local moments in Table  \ref{moments} we find that the isolated Ni $1a$ moment is increased as expected, but at the same time the Ni $3c$ moments are quenched due to the covalent interaction with the central N atom.

\subsection{Exchange interaction and Curie temperatures}
To assess the validity of the KKR calculations, we start by comparing the KKR results for the magnetic moments with the FLAPW results, which are taken as the reference values. The KKR results are in overall good agreement with the FLAPW results, with the largest discrepancies being present for Cr$_4$N and Ni$_4$N. In these two cases, an underestimation of the 3c magnetic moment is mostly responsible for the deviation from the FLAPW results. The agreement between the KKR results and the reference FLAPW calculation is sufficiently good to use the KKR ground states as starting points for the calculations of exchange interactions and Curie temperatures.

The site- and distance-resolved pair exchange parameters $J_{ij}$ are dominated by direct interactions within a distance of one unit cell. At larger distances, RKKY-like interactions come into play and decay approximately as $1/r^3$. To make the results more accessible, two additional representations are given in Figure \ref{fig:Jij_results}: the distance dependent mean-field Curie temperatures and graphical representations of the coupling matrices $J_0^{\mu\nu}$. The former contain all interactions within a sphere of given radius that are used to construct the mean field equation \ref{eq:mfa}. In the latter case, the matrix represents the total interaction between each pair of crystallographic sites with indices $\mu$ and $\nu$.

At short interaction distances, the following trends can be identified for the interactions. Each $1a$ site has twelve $3c$ sites as nearest neighbors at distance $r = a/\sqrt{2}$. The direct nearest-neighbor interaction between $1a$ and $3c$ sites is antiferromagnetic for Cr$_4$N and Mn$_4$N, whereas it is ferromagnetic in the other cases. These interactions are responsible for the ferrimagnetic ground states of Cr$_4$N and Mn$_4$N. Each $3c$ site has four $1a$ sites and eight $3c$ sites at distance $r = a / \sqrt{2}$ as nearest neighbors, ignoring the two nitrogen atoms at $r = a/2$. With the exception of Mn$_4$N, this nearest-neighbor $3c - 3c$ interaction is always ferromagnetic. The exchange between $3c$ sites through the cube center nitrogen atom, i.e. the nearest $3c$ intra-sublattice interaction at $r = a$, is always ferromagnetic. This situation is clearly different from, e.g., MnO, where a similar octahedron of Mn$^{2+}$ ions surrounds an O$^{2-}$ ion. In that case, opposing Mn ions are coupled antiferromagnetically via superexchange through the oxygen ion. The exchange between $1a$ sites through the central nitrogen atom, i.e. the $1a$ intra-sublattice interaction at $r = \sqrt{3}a$, seems to oscillate with the transition metal valence electron count, being ferromagnetic for even and antiferromagnetic for odd valence electron numbers.

Contrary to na\"{\i}ve expectation, the coupling matrices are very different for the five tetrametal nitrides. The coupling matrices of Cr$_4$N and Ni$_4$N can be seen as prototypical for simple ferri- and ferromagnets, respectively. In Cr$_4$N, the total intra-sublattice interactions (i.e. the interaction of atoms within one sublattice) and the total inter-sublattice interaction between crystallographically equivalent $3c$ atoms are all ferromagnetic, whereas the $1a-3c$ interaction is antiferromagnetic giving rise to the ferrimagnetic ground state. In Ni$_4$N, the total couplings are ferromagnetic, rendering Ni$_4$N a rather simple ferromagnet.

In the intermediate cases Mn$_4$N, Fe$_4$N, and Co$_4$N the situation is more complicated. In Mn$_4$N, the interactions are dominated by the $1a$ site, with the total intra-sublattice interaction being ferromagnetic and the total $1a-3c$ inter-sublattice interaction being antiferromagnetic. The total $3c$ intra-sublattice interactions are essentially zero and the total $3c$ inter-sublattice interaction is weakly antiferromagnetic. This has already been pointed out in an earlier theoretical study based on total energy calculations \cite{Uhl1997}. In Co$_4$N most coupling elements are ferromagnetic up to the $1a$ intra-sublattice interaction, which somewhat surprisingly turns out to be weakly antiferromagnetic. The most remarkable coupling matrix is that of Fe$_4$N, where all interactions sum up to ferromagnetic coupling with the exception of the $3c$ intra-sublattice interactions, which are strongly antiferromagnetic. As can be seen from the $T_\mathrm{C}^\mathrm{MFA}(r)$ graph, a major negative contribution from the $3c-3c$ interaction sets in at $r = \sqrt{2}a$ and leads to a drastic reduction of the MFA Curie temperature. The reduction was also confirmed by Monte Carlo calculations with a restricted interaction range. This peculiar interaction does in fact limit the Curie temperature in Fe$_4$N, which would otherwise potentially be a few hundred K higher.

The mean field Curie temperatures $T_\mathrm{C}^\mathrm{MFA}$ overestimate the experimentally known Curie temperatures by 15 - 30\%. In contrast, the Monte Carlo Curie temperatures $T_\mathrm{C}^\mathrm{MC}$ underestimate the experimental values by about 3\% in Mn$_4$N and Ni$_4$N and by 13\% in Fe$_4$N. Still, the overall agreement with the experimental results can be considered excellent. For Cr$_4$N we predict the Curie temperature to be close to room temperature and Co$_4$N is expected to have the highest Curie temperature among all the tetrametal nitrides. Surprisingly, the Curie temperature of Ni$_4$N is very close to the experimental value, whereas the calculated Curie temperature of fcc Ni (348\,K) at the experimental lattice constant is much lower than the experimentally observed value of 633\,K. This can be interpreted as a localization effect of the Ni $1a$ moment, which allows to apply the Heisenberg model in contrast to fcc Ni which has largely itinerant character. By reducing the coupling matrix of Ni$_4$N to the $1a-1a$ interaction, the MFA Curie temperature is reduced by only 11\%, indicating that the intra-sublattice interactions of  the localized moment on the $1a$ site are responsible to a large part for the Curie temperature of Ni$_4$N, making the Heisenberg model valid in this case. Finally it should be noted that no simple trend relating the total moment, absolute moment (sum of the absolute local moments) or any of the local moments to the Curie temperature can be identified.

\begin{figure}[t]
\centering\includegraphics[width=8.6cm]{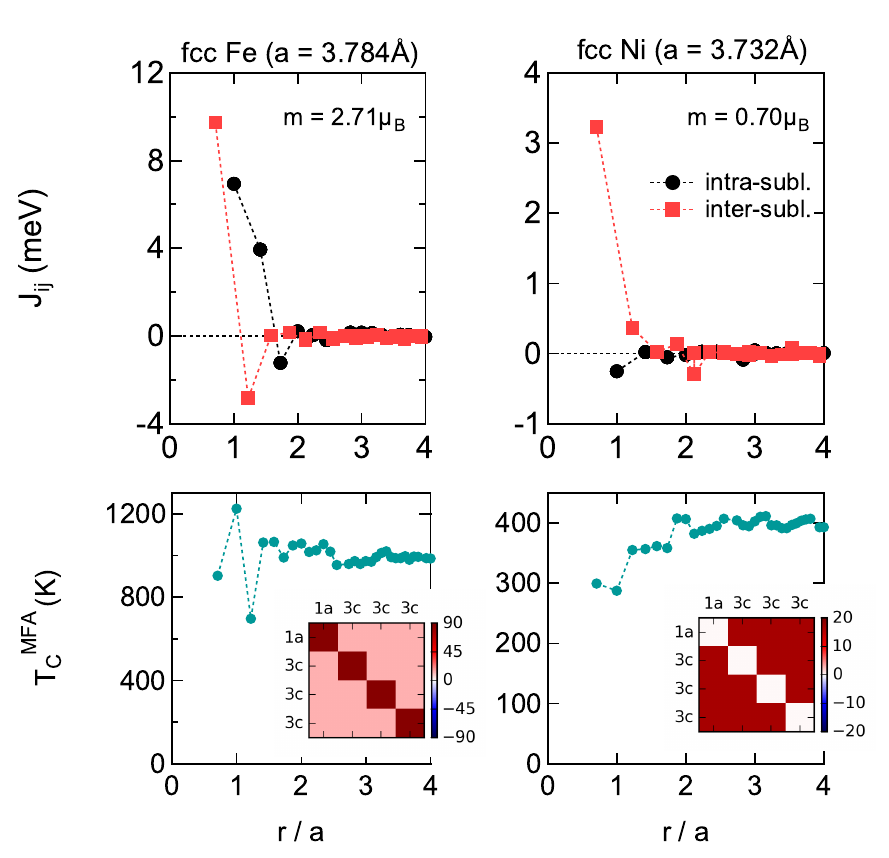}
\caption{\label{fig:Jij_fcc}Top row: Heisenberg exchange parameters as a function of interatomic distance in expanded fcc metals. Bottom row: Mean-field estimate of the Curie temperature as a function of interatomic distances included in the construction of $J_0^{\mu\nu}$. Insets: Graphical representation of the coupling matrices $J_0^{\mu \nu}$. Note the scale changes between different compounds. The Monte Carlo Curie temperature are $T_\mathrm{C}^\mathrm{MC}(\mathrm{Fe}) = 750$\,K and $T_\mathrm{C}^\mathrm{MC}(\mathrm{Ni}) = 337$\,K.}
\end{figure}

\subsection{Comparison with expanded fcc metals without nitrogen}
To study the role of the nitrogen atom on the exchange coupling, the Heisenberg parameters were calculated for fcc Fe and fcc Ni at the lattice constants of Fe$_4$N and Ni$_4$N, respectively. The results are presented in Figure \ref{fig:Jij_fcc}. $1a$ and $3c$ site labels are used equivalent to the tetrametal nitride case, however this distinction is somewhat artificial because these sites are crystallographically equivalent without the central nitrogen atom. In both cases, enhanced magnetic moments with respect to bcc Fe and fcc Ni at the experimental lattice constant are observed. Remarkably, the magnetic moments are also enlarged with respect to the nitrides at the same lattice constant, indicating that the covalent interaction  with the cube centered nitrogen quenches the magnetic moment of the face centered transition metal atoms. The magnetic moment of expanded fcc Fe is found to be $10.84\,\mu_\mathrm{B}$\,/\,cell in the KKR calculation, and $2.8\,\mu_\mathrm{B}$\,/\,cell is found for expanded fcc Ni. This observation may provide an alternative explanation for the broad range of magnetic moments observed for Fe$_4$N: at reduced nitrogen content, the magnetic moment is increased, whereas additionally incorporated nitrogen at the Wyckoff $3d$ interstices would further reduce the moment. Similarly, disorder of nitrogen between $3c$ and $3d$ positions could increase or decrease the magnetic moment, depending on the degree of disorder.

The exchange parameters of the expanded fcc metals are very different from the nitrides. This is evident at the first glance by comparing the representations of the coupling matrices in Figures \ref{fig:Jij_results} and \ref{fig:Jij_fcc}. In expanded fcc Fe, the positive intra-sublattice interactions dominate the total interactions, whereas in Fe$_4$N the largest contribution comes from the $1a - 3c$ interaction. In contrast, the total intra-$3c$ interaction is antiferromagnetic in Fe$_4$N, whereas it is ferromagnetic in expanded fcc Fe. However, the Monte Carlo Curie temperature $T_\mathrm{C}^\mathrm{MC}(\mathrm{Fe}) = 750$\,K of the expanded fcc Fe is very similar to that of Fe$_4$N.

In fcc Ni the inter-sublattice interactions dominate the total exchange interactions, whereas the intra-sublattice interactions are comparatively weak. This is again very different from the nitride, where the $1a - 1a$ interaction dominates the exchange interactions. The Monte Carlo Curie temperature of expanded fcc Ni ($T_\mathrm{C}^\mathrm{MC}(\mathrm{Ni}) = 337$\,K) is very close to the value of fcc Ni at the experimental lattice constant, whereas the value of Ni$_4$N is roughly three times smaller. This finding underlines that the tetrametal nitrides cannot be simply interpreted as fcc metals with enhanced lattice constant. Instead, the covalent interaction with the body-centered nitrogen atom has large impact on the electronic structure of the compounds.

\section{Summary}
The exchange interactions and Curie temperatures of the five possible magnetic tetrametal nitrides were calculated and very good agreement with experimental data was obtained. It was shown that the exchange interactions are neither intuitive nor do they vary systematically across the series of compounds ranging from Cr$_4$N to Ni$_4$N. Interestingly, the Curie temperature of Fe$_4$N is limited by the surprising presence of a strong antiferromagnetic intra-sublattice interaction between the cube corner atoms. A comparison between Ni$_4$N and fcc Ni showed that the tetrametal nitrides cannot be seen as fcc metals with expanded lattice constant.

\end{document}